\documentstyle[aps,psfig]{revtex}

\def\x{\vec{x}}

\def\bfg{\begin{figure}}
\def\efg{\end{figure}}
\def\be{\begin{equation}}
\def\ee{\end{equation}}
\def\REF#1{\cite{#1}}

\begin{document}
\pssilent
\title{Locomotion and proliferation of glioblastoma cells {\it in vitro}:
statistical evaluation of videomicroscopic observations} 

\maketitle

\vbox to 1cm {} 

Bal\'azs  Heged\H us$^1$  M.Sc., ~  Andr\'as Czir\'ok$^1$   M.Sc.,
 ~ Ilona Fazekas$^2$   M.D. Ph.D., 

Tam\'as B\'abel$^2$  M.D., ~ Em\'\i lia Madar\'asz $^3$   M.Sc. Ph.D.   \& 
  
Tam\'as Vicsek$^1$  M.Sc. Ph.D.  University Professor 

\bigskip

$^1$  Department of Biological Physics, E\"otv\"os University, Budapest, 
 Hungary
 
$^2$ National Institute of Neurosurgery, Budapest, 
 Hungary 

$^3$ Institute of Experimental Medicine, Hungarian Academy of Sciences,  
Budapest, Hungary

\bigskip

\begin{tabular}{rl}
{\bf Corresponding and reprint author:} &
Bal\'azs  Heged\H us\\
& Department of Biological Physics\\
& E\"otv\"os University\\
& H-1117 Budapest\\
& P\'azm\'any P\'eter s\'et\'any 1A\\
& Hungary \\
& Tel: (36)-1-3722795\\
& Fax: (36)-1-3722757\\
& e-mail: hebal@angel.elte.hu
\end{tabular}

\bigskip

\begin{tabular}{rl}
{\bf Supports:} &
Gedeon Richter Pharmaceutical Company (Richter 216/97/KK)\\
& Ministry of Education (FKFP 0203/1997)\\
& Hungarian Science Foundation (OTKA T25719, OTKA T019299)\\
\end{tabular}

\vbox to .5cm {}

{\bf Running head}: Locomotion and proliferation of glioblastoma cells

\medskip



\section{abstract}
Long-term videomicroscopy and computer-aided statistical analysis 
were used to determine some characteristic parameters of {\it in vitro} 
cell motility and proliferation in three established cell lines derived 
from human {\it glioblastoma} tumors. 
Migration and proliferation activities were compared among the three cell lines
since these are two features of tumor cells that strongly influence the 
progression of cancer.

Cell proliferation in sub-confluent cultures were evaluated by calculating 
the growth rate of the number of cells and the distribution of the cell-cycle 
times in a given microscopic field. In these parameters no significant 
difference was observed among the cell lines regardless to the number of 
passages.

Studies on cell locomotion revealed strong fluctuations in time and 
exponential distribution of cell velocities. In spite of the fluctuations, 
both the distribution profile and the average velocity values were reproducibly 
characteristic to each cell line investigated. 

 The results on these dynamical parameters of cell locomotion were compared to 
pathological data obtained by traditional methods. The data indicate that the 
analysis of cell motility provides more specific information and is potentially
useful in diagnosis.



\section{Introduction}

In order to provide appropiate methods for the characterization of the large 
variety of human brain tumors, reproducible and quantitative models are needed.
Then, one can carry out a  characterization of the dynamics 
of tumor cells providing substantial information supplementing the results of 
such statical methods as histology or molecular genetics. 
For example, {\it in vitro} studies on the cell cycle and motility of tumor 
derived cells could provide relevant information for the diagnosis and the 
estimation of the progression of the cancer. 
The correlation between tumor malignancy and increased migration activity of 
glioma cells had been demonstrated by {\it in vivo} and {\it in vitro} 
experiments using rat brain xenografts \REF{Bern7,neu83} or radial dish assays 
\REF{rda,rev}. In addition, studies of cell proliferation using either 
proliferation-marker antibodies in tissue sections or flow cytometry of tumor-
derived cell cultures have also demonstrated that the growth fraction of the 
specimen often correlates with the malignancy of the tumor \REF{proli1,ki67}.

In several cases tumor recurrence is originated by a minor subpopulation
of highly migratory cells which were not removed surgically with the bulk
tumor. Thus, the detection of such subpopulations and the estimation of their
infiltration depth is essential for postoperational treatments.
However, fluctuations in locomotion or proliferation generate difficulties in 
the characterization of the heterogeneity of a tumor specimen; 
according to early studies a strong variation is present in the 
intermitotic times even within a single-cell derived population \REF{nat77}. 
Fluctuations dominate the motility of single cells as well \REF{dunn,mozgas1,cza1}.  
The classical methods assaying cell division frequency or cell motility cannot 
distinguish between a fluctuating but homogeneous population and an 
inhomogeneous population, thus their prognostic value is rather limited 
regarding heterogeneity \REF{ki67}. To reveal details about heterogeneity and 
subpopulations {\it in vitro} experiments and careful statistical analyses are 
required.
 
Simultaneous and automatized observation of the mitotic and
locomotory activity of a large number of cells  during a long time interval
can provide a direct way to overcome such difficulties.  

We describe here a novel long-term videomicroscopic system together with a
computer-aided statistical analysis to quantify the {\it in vitro} behavior of
cells. With this method we compared three established glioblastoma cell lines
derived from three patients.  Glioblastoma is a common and highly malignant
(WHO Grade IV.) tumor that develops in late adult life and generally located
in the cerebral hemispheres.  These tumors contain motile and invading cells,
frequently causing rapid recurrence after surgical resection \REF{Naz}.
In some cases the migration of these cells lead to the progress of the disease 
even without the formation of notable mass effect \REF{mass}.

While our studies revealed a rather similar proliferation activity in all
investigated cell lines, a reproducible and significant difference was found
in the locomotory activity of the cells. Due to the large number of
individually tracked cells the question on the homogeneity of cell motion and 
proliferation of the cell populations could be also addressed. 
By appropriate automatization the
complete analysis can be completed within 4 or 5 days after processing the
surgical specimen to one-cell suspension, providing a new tool for
diagnostical purposes. 


\section{ Materials and Methods }
 
\subsection{Diagnosis of human brain tumors}

 The surgical specimens were obtained during craniotomies for resection of 3 
hemispherical primary brain tumor. The bigger part of the specimen was 
evaluated by routine histopathology, including staining with hematoxylin and 
eosin and immunohistochemical staining for GFAP expression. Specimens were 
graded according to the WHO classifications \REF{who} at the Histopathological 
Department of National Institute of Neurosurgery.

\subsection{Establishment of human brain tumor cell lines}
 
 Part of the specimen was used for establishing the cell lines. The samples 
were washed by Minimal Essential Medium (Sigma) containing gentamycin (Chinoin)
and Fungizone (Gibco). After removing the vessels, the samples were minced and 
then triturated with a Pasteur-pipette. The primary tissue was seeded into 
Leighton culture flask on Bellco-slide or into Steriline culture flask in 
Dulbecco's MEM (Sigma) with 20\% FCS (Gibco). According to our procedure the 
cultures are not one-cell derived. The passages were performed with 0.25\% 
trypsin and cells transferred to Greiner tissue culture flasks. Cells were
maintained in DMEM (Sigma) with 10\% FCS, 40 $\mu$g/ml gentamycin and 
5 $\mu$g/ml Fungizone, at $37^o$C in a humidified 5\% CO$_2$ atmosphere.
 Cells were stored in 10\% DMSO (Sigma) DMEM solution in liquid nitrogen for 
later use.

\subsection{Immunocytochemical staining for GFAP in the cultures}

 The cultured cells' glial origin was indicated by the presence of glial 
fibrillary acidic protein (GFAP). Cells cultured on slides were fixed with 
Zamboni-solution for 60 minutes at room temperature. Then incubated as follows:
3 times 10 minutes in PBS (Phosphate Buffered Saline), 20 minutes 15\% human 
albumin in PBS (Human), 30 minutes with mouse anti-GFAP antibodies (High 
Performance, Biogenex) in a humidified chamber, 20 minutes biotin-conjugated 
anti-mouse immunoglobulin (Multilink Stravigen, Biogenex), 20 minutes 
peroxidase-conjugated streptavidin (Biogenex). Slides were developed with DAB 
(2,5 mg DAB+ 5 ml PBS+ 50{$\mu$}l 1\% H$_2$O$_2$) under microscopic control. 
Controls were performed without the primary antibodies.

\subsection{Long-term videomicroscopy}  
 
$10^5$ cells, counted in haemocytometer, were plated in 35 mm TC-dishes
(Greiner) with 2 ml medium (DMEM (Sigma) with 10\% FCS (Gibco)). Cell cultures 
were kept in a mini-incubator -- providing $37^o$C in a humidified 5\% CO$_2$ 
atmosphere -- attached to the powered stage of an inverted phase-contrast 
microscope (Zeiss Televal-1)(Fig 1).  Images of 3-6 neighbouring microscopic 
fields were taken in every 5 minutes during a 3-day long period, with 
10$\times$, 20$\times$ or 40$\times$ objectives using a CCD camera (JVC KY-F30B)
connected directly to the frame grabber card (Matrox Meteor, Matrox Electronic 
Systems LTD, Canada) of a PC (running under LINUX operating system).

\subsection{Cell positions and trajectories} 

To determine the position ($\vec{x_i}$) of the individual cells, the
geometrical center of each cell was tracked manually in every 4th image (i.e.,
in every 20 minutes in real time). The difference in the location of a given
cell was considered as a migratory segment. Trajectories were constructed from
these segments.

\subsection{Duplication time}

Using the database of cell positions, the total number of cells in a given
microscopic field ($n$) was determined.  The growth of cell number was found
to be approximately exponential in each investigated culture, i.e., $n \sim
e^{\alpha t}$. Based on a least square-fit of $\alpha$, the duplication time,
$\tau$, was calculated as $\tau=(\ln2)/\alpha$.

\subsection{Cell-cycle time and non-proliferating cells} 

All cell divisions during the observation period were identified resulting 
a data base containing the corresponding mother and daughter cells. The cell
cycle length ($\tau_0$) of a given cell was then determined as the time
elapsed between two consecutive cell divisons.  

We denote by $\nu$ the rate at which non-proliferating cells appear in the 
culture. It was calculated in two steps. First we determined $N$, the number 
of proliferating cells by the criteria that both their birth and subsequent 
mitosis was recorded and their birth occured more than 30 hours before the 
end of the recording. Then $N_*$, the number of suspected non-dividing cells 
was calculated. These cells were selected using the following criteria: (i) 
lack of observed mitosis during the entire recording
period, and (ii) being tracked for at least 30 hours following their birth.
Then, $\nu$ was calculated as $\nu=N_*/(N_*+N)$.

Note, that there are two sources of systematic errors in this procedure.  On
one hand, when calculating $N_*$, cells may be included which do divide, but
their cell cycle length is longer than the time period during the culture was
observed. On the other hand, we did not include in $N_*$ cells which do not
divide, but we could not track them for longer than 30 hours because they left
the field of observation. Taking into account that during a 30-hour period
approximately $10\%$ of the cells migrated out of the observed area, and
(based on the empirical distribution of the cell-cycle times) $\approx 20\%$
of the dividing cells have longer cell cycle than 30h, we estimate the
relative error of $\nu$ to be $10\%$.

Assuming that after each division a $\nu$th portion of daughter cells become
unable to further proliferate, the size of the cell population ($n$) after the
$k$th complete set of cell division -- denoted as $n(k\tau_0)$ -- is given by 
\begin{equation}
\log n(k\tau_0)\approx k\log 2(1-\nu) + const.
\end{equation}
Thus based on this simple calculation, the relation
\begin{equation}
\tau_0\log2=\tau\log2(1-\nu)
\label{zizi2}
\end{equation}
is expected to hold among the independently measurable parameters
$\tau$,$\tau_0$ and $\nu$.

\subsection{Cellular velocities}

The velocity, $v_k(t)$, of a given cell $k$ was calculated as the
translocation of its geometrical center during a given time interval, $\Delta
t$, i.e.,
\be
v_k(t)=\left\vert{\x_k(t+\Delta t)-\x_k(t)\over\Delta t}\right\vert.
\ee
$\Delta t$ was chosen to be $1$h, since during this time interval the 
displacement of the cells were typically larger than the cell size, thus the 
relative error of $v_k(t)$ resulting from the tracking procedure was reduced. 
Nevertheless, during one hour (as our data show) the fluctuations in cell 
motility were not yet averaged out.

To characterize such a fluctuating quantity as the cell locomotion activity we
determined the $F(v)$ cumulative distribution function. $F(v)$ gives the
probability of the event that the velocity of a randomly selected cell at a 
given time is less than $v$. Since the relative error of the small cell
displacements is high due to the manual tracking procedure, velocities less
than 5 $\mu$m/h were discarded in the analyses.  The $F(v)$ empirical
distribution function was fitted by an exponential distribution \REF{cza1} 
as 
\be 
F(v)=1-\exp(-v/v_0), 
\label{zizi}
\ee 
where $v_o$ is a fitting parameter being equal to the mean velocity if
(\ref{zizi}) is exact.

As another way to characterize the motility of the cells, the empirical
average velocity of the culture, $V$, was calculated as
\be
V={\sum_k\sum_i{v_k(t_i)}\over\sum_k\sum_i{\delta_k(t_i)}},
\ee
where $\delta_k(t_i)$ is 1, if the $k$th cell is in the visual field in the
$t_i$ moment, otherwise 0.


\section{results}

\subsection{Histological description of tumors and cell lines}
 
The main features of the cell lines are given in Table 1.  The routine
histology of the tumor specimens concerning WHO classification revealed
glioblastoma. All three sections showed hypercellularity, pleomorphism,
vascular proliferation and necrosis. The HA and HC specimens (detailed in
Table 1) were classified as {\it glioblastoma multiforme}. In the HB specimen
some areas were dominated by multinucleated giant cells and the invasion of
lymphoid cells was observed. Thus HB was classified as giant cell variant of
glioblastoma.

The cultures were morphologically characterized according to Bigner
\REF{big}. Usually the HA-derived cells had two or three processes and the
formation of filopodia was quite frequent.  The HB culture's cells were mostly
elongated, had more processes and the formation of filopodia and lamellipodia
with ruffling edges was exceedingly intensive. Both morphology was described
as fibroblastic.  In contrast, even the subconfluent HC cultures showed typical
epithelial morphology with the general absence of cellular processes. The cells
formed preferentially a monolayer, though cells could crawl over each other.
During cultivation multinucleated monstrocells were observed in all the three
lines. On the time-lapse records cell divisions into three daughter cells were
also observed occasionally. In most of these cases the resulting daughter
cells seemed to perform normal mitoses later on.

\subsection{ Cell proliferation }
 
The number of cells in a given microscopic field grew exponentially after
an initial lag phase of 15-20 hours (Fig 2). Based on these data the cell
duplication times were determined (Tab 2). No significant differences were
found in the duplication time of the various cell lines: each culture could be
characterized by $\tau$= 38$\pm$4 hour. 

The cell number in a given microscopic field changed because of cell divisions
and the migration of the cells. According to our observations the net current
of migration was approximately zero, so the increase of cell number was
basically determined by the ratio of dividing cells and the duration of their
cell-cycle.  

Continuous observation of the cultures allowed to determine the cell-cycle
time ($\tau_0$) of individual cells as well. The empirical distribution 
functions of the cell cycle lengths were similar in all the three cell lines 
investigated (Fig 3). In fact, based on this amount of data the distributions 
could not be distinguished based on a Wilcoxon test (with p$<$0.05). The 
average cell-cycle time was found to be 25.6$\pm$6.2 hours.  The difference 
between the average cell-cycle time and the duplication time of the cell 
population can be explained by the presence of non-dividing cells. The 
proportion of these cells, $\nu$, was found to be $32\pm 5\%$  (Table 2).
According to relation (\ref{zizi2}) this proportion was determined by average 
cell-cycle time ($\tau_0$) and duplication time ($\tau$) as 34\%.  
Thus, relation (\ref{zizi2}) holds within error for the rate of arising of 
non-proliferating cells, average cell-cycle time ($\tau_0$) and duplication 
time ($\tau$).

To investigate the role of hereditary factors in the determination of
cell-cycle time the correlation between parent and daughter cells was
studied (Fig 4). The scattering of data indicates uncorrelated variables,
it means that the duration of cell-cycle is not influenced by the parent cell's cell-cycle time.

\subsection{ Cell locomotion} 
 
The dynamics of cell shape can be qualitatively observed and characterized
using the long term time-lapse records(Fig 5). The videomicroscopic images
showed the rich dynamics of the formation of lamellipodia and filopodia. The
intensity of ruffling edge formation correlated with the motility of cells \REF
{mozgas1}. As expected, the epithelial \REF{big} HC cells displayed the lowest 
motility and the HB cells were more motile than the other fibroblastic \REF{big}
HA cell line.  

The analysis of the pathways of individual cells revealed that in
subconfluent cultures the direction of cell movement is random and disordered
(Fig 6a).  However, at high cell densities the HB cell line displayed a rather
ordered collective migration (Fig 6b). 

In all of the three cell lines the cell velocities were highly fluctuating as 
demonstrated in Fig 7. 

The $F(v)$ cumulative distribution functions are shown in Fig 8. Both the
distribution functions and the average velocities of the culture ($V$ and
$v_0$) were different among the three cell lines but within error reproducible 
for each individual cell line (Table 3).

\subsection{Population homogeneity}

To understand the relation between velocity distribution function and the 
homogeneity of locomotory activity of the cell population further statistical 
analysis is required.   
The velocity distributions described by (\ref{zizi}) were found for relatively
large ($\approx 100$ cells) populations. On the level of individual cells the
exponential behaviour (\ref{zizi}) can be interpreted in two ways: (i) The
culture is inhomogeneous, i.e., slower and faster cells can be distinguished
on the bases of well preserved phenotypic properties.  In this case the
exponential $F(v)$ distribution can reflect the ratio of the slow and fast
cells in the culture, while the velocity fluctuations of the individual cells
can show an arbitrary distribution.  (ii) If the culture is homogeneous, then
almost all cells exhibit the same distribution of velocity fluctuations, i.e.,
$F_i(v)\approx F(v)$ holds for each cell $i$, where $F_i(v)$ denotes the
velocity distribution function of the cell $i$. In this scenario the
time-averaged velocity $\langle v_i(t) \rangle$ of each cell would be the same
if we could calculate the time averages over an infinitely long time. Since
the time averages are calculated over a finite time $T$ only, for the
distribution of the average velocities we can expect a Gamma distribution with
a parameter $s=T/t_0$, where the correlation time of the process is denoted by
$t_0$ \cite{cza1}.

We used the distribution function of the average velocity of individual cells
(averaged over the entire observation period) to characterize the
inhomogeneity within the cell lines. In all cases the average velocity of the
cells showed a gamma distribution indicating the homogenity of the cell
population (Fig. 9.).  However, if applying the same kind of analysis on a
cell position data base in which cells from different cell lines were mixed, a
significant deviation from the gamma distribution can be observed (shown in the insert of Fig. 9.).

\section{discussion}

The comparison of classical morphological data and the dynamical properties in 
the three {\it glioblastoma} cell lines gave some unexpected results.
In spite of the different origin and different passage-levels of the 
investigated lines significant alterations in the {\it in vitro} proliferating 
potential were not revealed. Both the cell-cycle time and the percentage of 
dividing cells under the same culture conditions were found equal in the three 
different {\it glioblastoma} lines. The cell-cycle time measured in our
cultures did not differ from the doubling time of normal primary fibroblasts 
from human skin \REF {nature98}. The variability of the cell-cycle time of 
individual cells seems to be independent from inheritable factors and might be 
best explained by the stochastic activity of the cells. 

The experimental verification of Eq. (\ref{zizi2}) suggests that the
non-dividing cells did not belong to an initial subpopulation in the culture
but arose continuously during proliferation. The considerable proportion of
these cells may indicate that defective mitoses are abundant in the cell 
culture. The cause of this phenomenon can be the genetical instability of tumor
cells which has been recently demonstrated \REF {nature98}.

 Although the {\it in vitro} cell motility is a highly stochastic process 
regarding the direction of the movement or the fluctuation of the velocity of 
an individual cell, the locomotory activity of the cell lines have some 
quantitative properties that can be interdependent with the morphology of the
cultured cells. The cell lines with fibroblastic morphology performed higher 
locomotory activity. Reproducible and significant difference was found in the 
locomotion of the cells among the three investigated {\it glioblastoma} cell 
lines. 

The HB case can illustrate that classical studies are not always satisfying in 
the forecast of the disease course. In spite of the diagnosis as a giant cell 
variant, which is commonly known as a less malignant type of glioblastoma, and 
the powerful presence of immune response, the tumor was fatally renewed within 
4 month. It can be supposed that the fast and strong recurrence was due to the 
{\it highly motile cells} that invaded the surrounding tissue and were not 
removed surgically.

Our data were obtained on established cell lines after multiple passages, in 
this way these cultures may not reflect entirely the properties of the cells of
the investigated tumors. With similar preparations the evaluation can be 
performed on primary cultures with no passages as well.
Investigations of {\it in vitro} dynamic behavior of primary cell cultures
from surgical specimens can complement classical diagnostic methods.  Although
during culturing an inevitable selection takes place, some of the initial
features or even heterogenity of the tumor tissue may be reflected in the cell
cultures.  If such analysis can be carried out within a short period of time,
it can help to choose the most appropriate postoperative therapy in each
individual case.

\section{Acknowledgements} 

We are indebted to Fel\'\i cia Slowik for the diagnosis.
We thank Zolt\'an Csah\'ok and Ott\'o Haiman for the technical support 
 and Antaln\'e Kerekes for the maintainance of the cultures. 


\def\PRL {{ Phys. Rev. Lett. }}



Figure legends:

\begin{figure}
\caption{Schematic representation of the long-term videomicroscopy system.
The incubator, placed on the inverted microscope's stage, holds the TC-dish.
The microscope stage and the adequate cell culture conditions (temperature, 
atmosphere) are controlled by the computer.
The CCD camera's images are transmitted to the computer, where the digitalized
images can be processed and analysed or recorded on videotape.}

\label{fig1}
\end{figure}

\begin{figure}
\caption{The growth of the normalized cell number in a given visual field.
The y-axis is logarithmic, and the actual cell numbers were normalized by the 
$N_0$ coefficient of the exponential fit $N\sim N_0e^{\alpha t}$. The growth 
rate is similar in all the cell cultures investigated. The solid line shows 
the fitted exponential growth with the average duplication time ($\tau$= 
38$\pm$4).}
\label{cellnum}
\end{figure}

\begin{figure}
\caption{The distribution of the cell-cycle time ($\tau_0$). The distribution 
functions obtained from the three cell lines could not be distinguished
statistically. Thus the proliferating cells in any of the investigated
cultures divided with the similar cell-cycle time.}
\label{cycle}
\end{figure}

\begin{figure}
\caption{ The daughter cells' cell-cycle time ($\tau_0 (d)$) versus the parent 
cells' cell-cycle time ($\tau_0 (p)$). The scattering of the data points means 
that the cell-cycle times in the subsequent generations do not correlate.}
\label{apafia}
\end{figure}

\begin{figure}
\caption{ Snapshots from the recordings of the various cell lines. Despite the
same diagnosis, morphological differences can be observed. The approximately
epithelial morphology of HC {\it (c)} cells can be compared to the fibroblastic 
characteristics of HA {\it (a)} and HB {\it (b)} cells.
Field of view is 740 $\mu$m $\times$ 560 $\mu$m.}
\label{kep}
\end{figure}

\begin{figure}
\caption{ Cellular trajectories in the HB3 measurement. {\it (a)} During the 
first 24 hours the cells perform a persistent random walk. {\it (b)} In the
second 24 hours, at a higher cell density, the cells show a spatially ordered 
migration with increased persistence length. Field of view is 740 $\mu$m * 560 
$\mu$m.}
\label{path}
\end{figure}

\begin{figure}
\caption{Cell velocities fluctuate in time. In the figure 3 representative 
cells' velocities are plotted vs time for each cell line. For better 
visualisaton the consecutive curves are shifted vertically by 50$\mu$m/h.}
\label{fluct}
\end{figure}

\begin{figure}
\caption{ The distribution function $G(v)=1-F(v)$ of the cell velocities
in the investigated cultures on linear-logarithmic plot. The linearity of the 
curves indicate that the  cell velocities can be described by an exponential 
distribution. The graph also shows that $G(v)$ function reproducibly 
distinguish the three different cell lines.}
\label{veldis}
\end{figure}

\begin{figure}
\caption{ Distribution functions of the average velocities of individual 
cells. The average was calculated over the entire measurement period and the 
functions are normalized by the average velocities of the cultures. The curves 
are well fitted by the calculated gamma distribution suggesting a 
phenotypically homogeneous population with respect to the locomotory activity.
In the inserted graph the mixed curve -- data from two different cell lines -- 
shows a significant difference from the calculated gamma function.} 
\label{gamma} 
\end{figure}

\newpage

\begin{table}
\caption{The most important data of the cell cultures analyzed}
\begin{tabular}{|c|c|c|c|c|c|c|}
\hline
measurement&HA1&HA2&HB3&HB4&HC0&HC2\\
\hline
date of &1997&1997&1996&1996&1995&1995\\
operation&March&March&October&October&February&February\\
\hline
age and sex& 58 y & 58 y & 51 y & 51 y & 61 y & 61 y\\
of patient& f & f & m & m & f & f \\
\hline
passage& 7 & 8 & 45 & 47 & 103 & 107\\
\hline
area of&&&&&&\\
observed& 1.14&0.41&0.41& 1.14& 1.14&0.41\\
field [mm]$^2$&&&&&&\\
\hline
initial cell&55&27&61&110&93&33\\ 
number&&&&&&\\ 
\hline
initial cell& 48& 65&148&96&82&80\\
density [mm]$^{-2}$&&&&&&\\
\hline
\end{tabular}
\end{table}

\begin{table}
\caption{The data on proliferation in the six measurements}
\begin{tabular}{|l|c|c|c|c|c|c|c|}
\hline
measurement&HA1~~~~&HA2~~~~&HB3~~~~&HB4~~~~&HC0~~~~&HC2~~~~& average~~~~\\
\hline
average cell-cycle&&&&&&&\\
time and its standard&24.2 $\pm$ 6.6 &27.9 $\pm$ 5.8 & n.a. & 24.7 $\pm$ 5.9 & n. a. & 25.9 $\pm$ 5.9& 25.6 $\pm$ 6.2\\
deviation [h]&&&&&&&\\
\hline
duplication time[h]&39,6&33,8&42,0&40,77&43.1&33,8&$38\pm4$\\
\hline
rate of non-&&&&&&&\\
proliferating cells& 39\% & n.a. & n.a. & 24\% & n.a. & 30\% & $32\pm5\%$\\
\hline
\end{tabular}
\end{table}

\begin{table}
\caption{The average velocities in the six measurements}
\begin{tabular}{|l|c|c|c|c|c|c|}
\hline
measurement& HA1~~~~& HA2~~~~& HB3~~~~& HB4~~~~& HC0~~~~& HC2~~~~\\
\hline
$V$ [$\mu$m/h]& 11,01 & 9,19 & 15,6 & 18,6 & 4,2 & 6,4 \\
\hline
$v_0$[$\mu$m/h]&11,6&11,1&15,6&17,2&6,5&7,0\\
\hline
error [$\mu$m/h]&1,2&1,1&1,6&1,7&0,7&0,7\\
\hline
\end{tabular}
\end{table}

\begin{figure}
\centerline{\psfig{figure=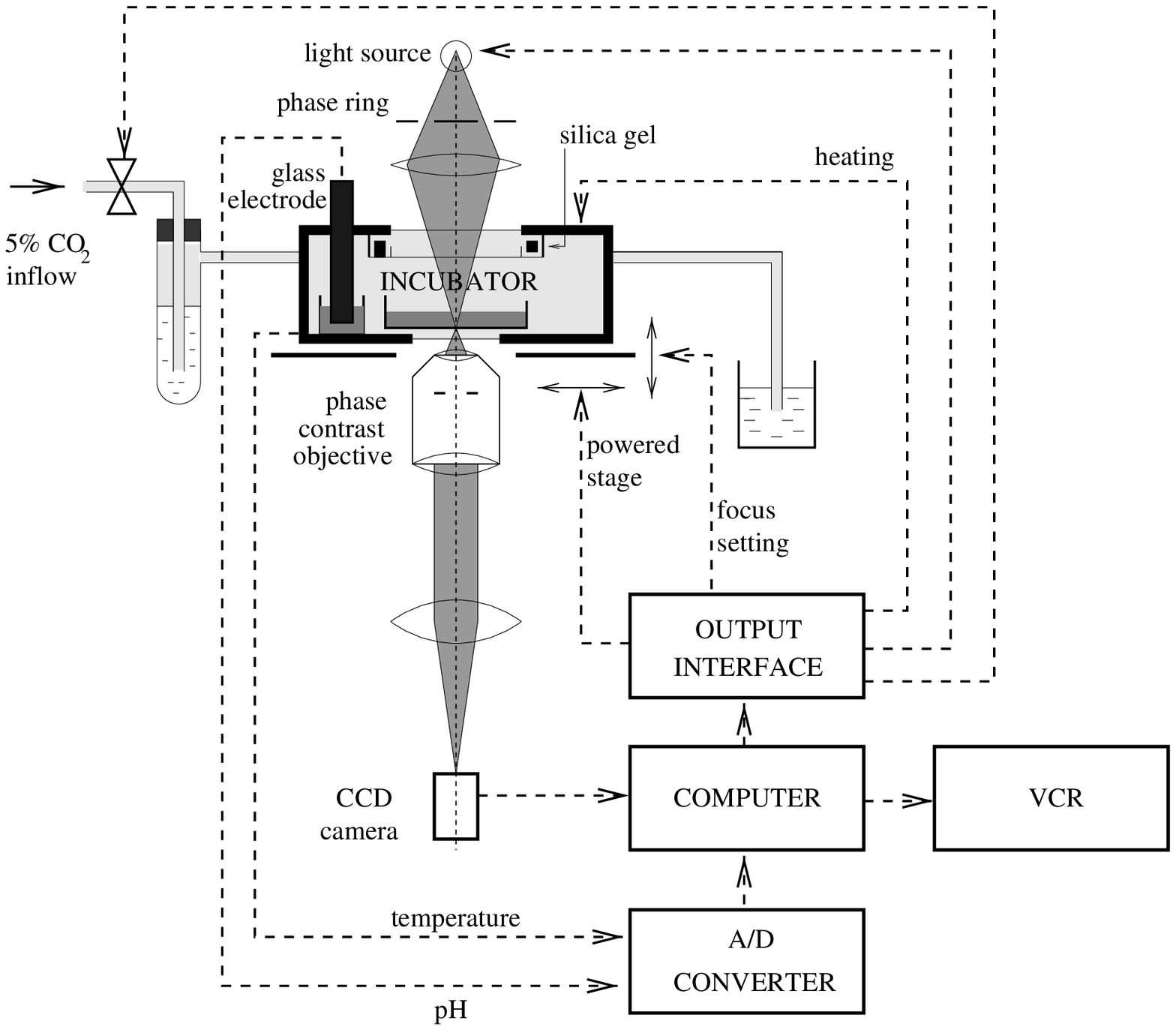,height=4.8in,angle=-90}}
\label{fig1x}
\end{figure}

\begin{figure}
\centerline{\psfig{figure=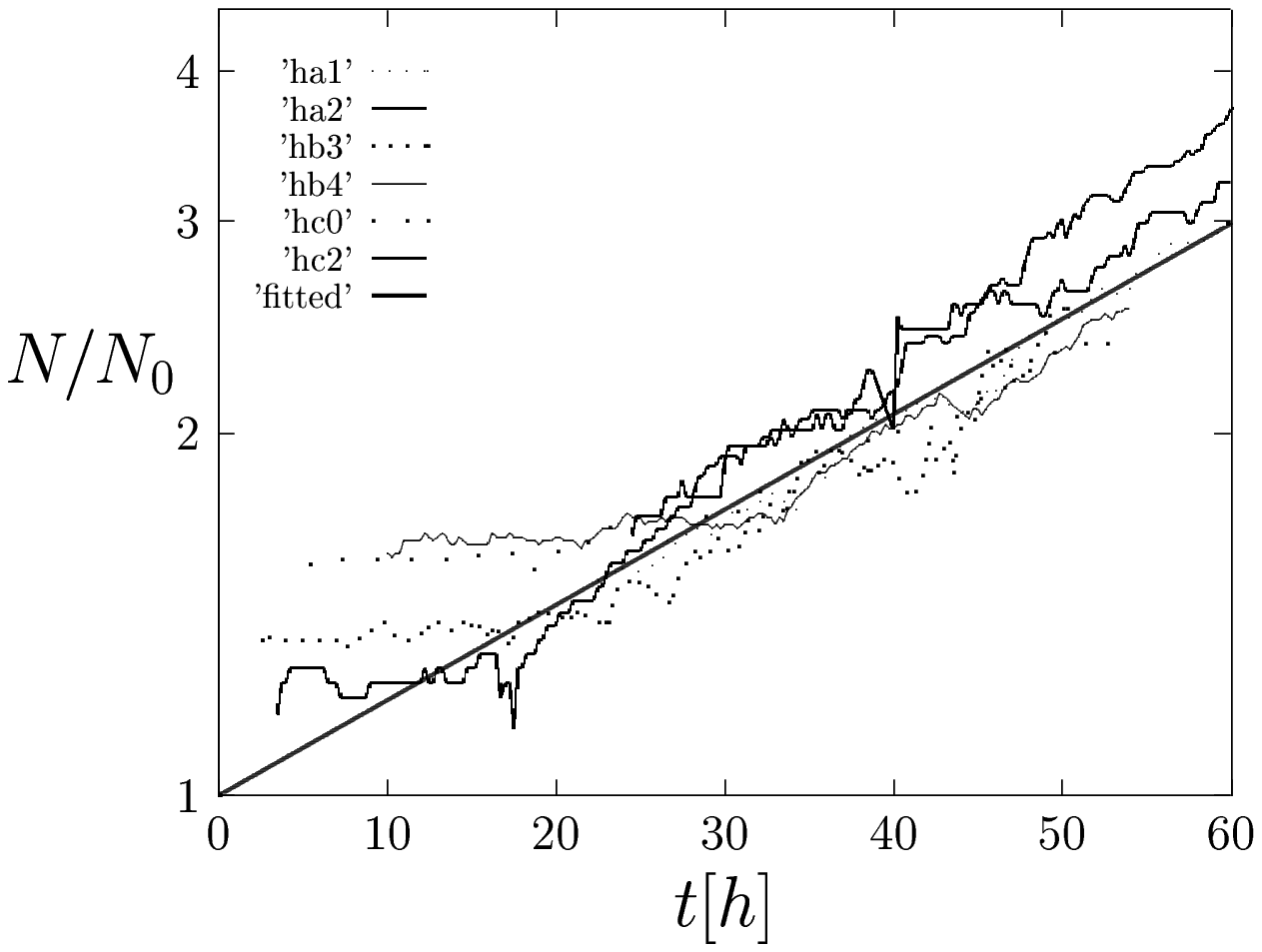}}
\label{cellnumx}
\end{figure}

\begin{figure}
\centerline{\psfig{figure=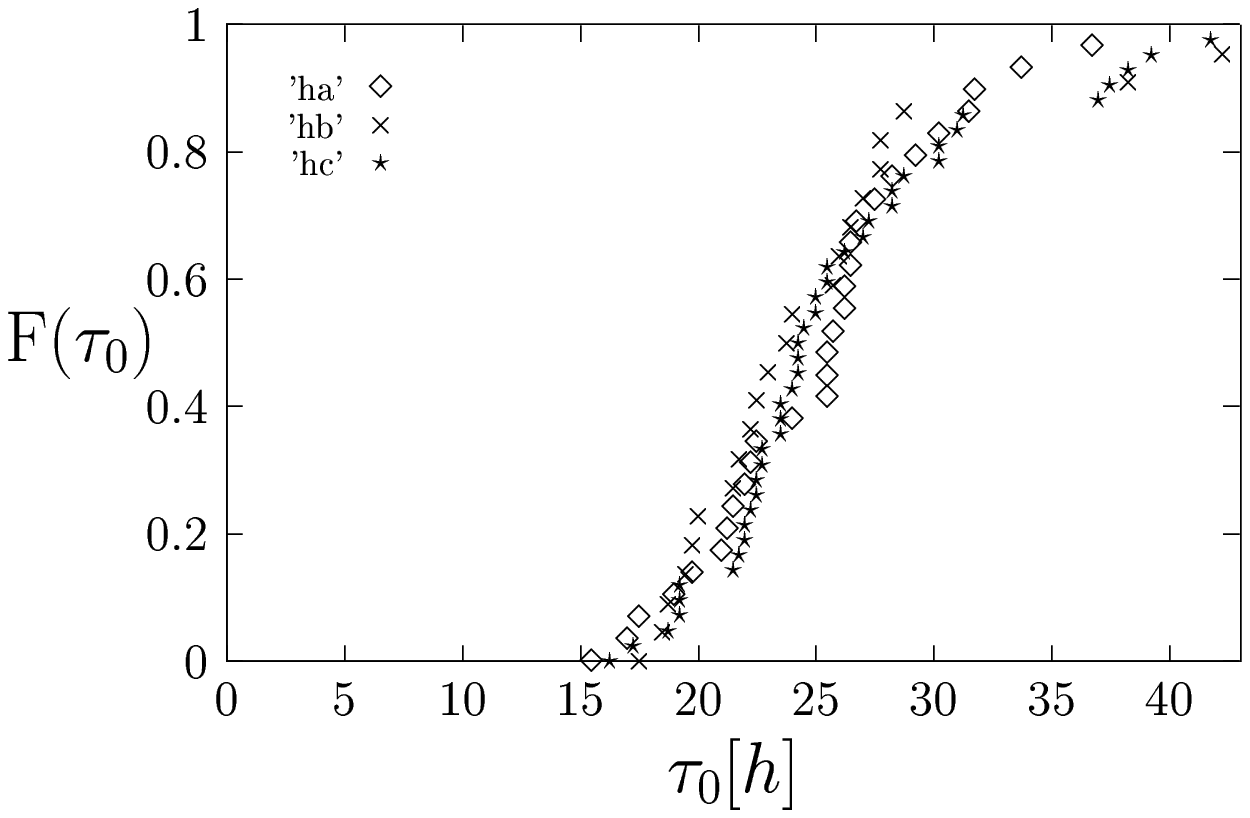}}
\label{cyclex}
\end{figure}

\begin{figure}
\centerline{\psfig{figure=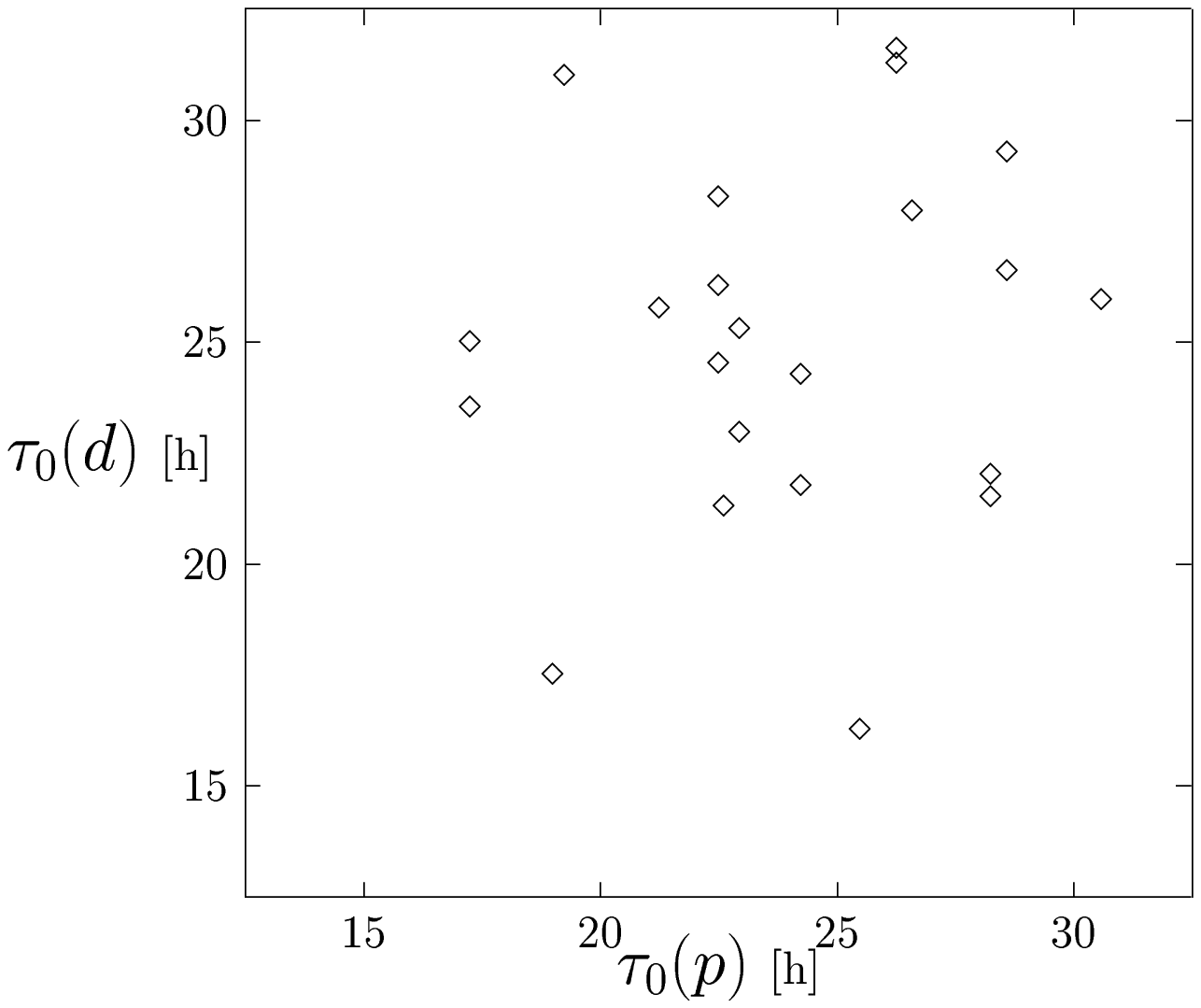}}
\label{apafiax}
\end{figure}

\begin{figure}
\centerline{\psfig{figure=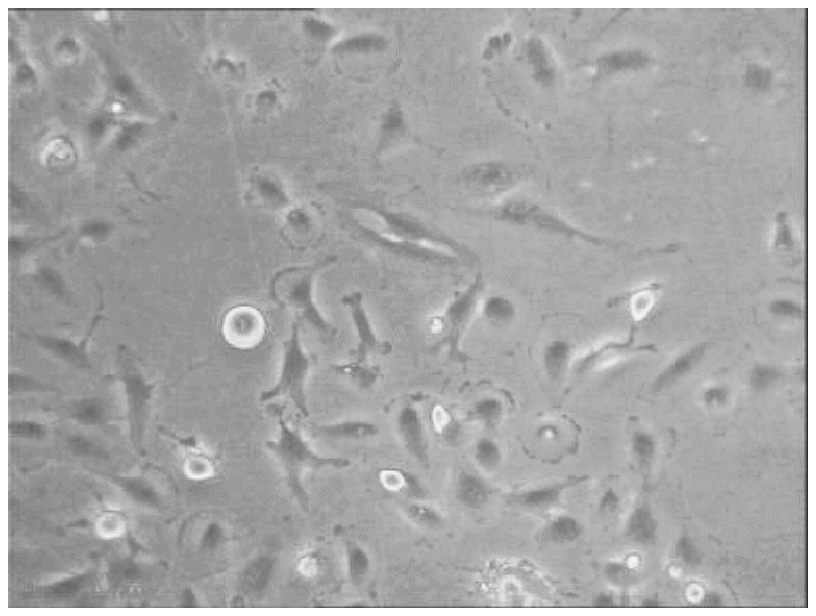,height=2.7in}}
\label{hc2x}
\end{figure}

\begin{figure}
\centerline{\psfig{figure=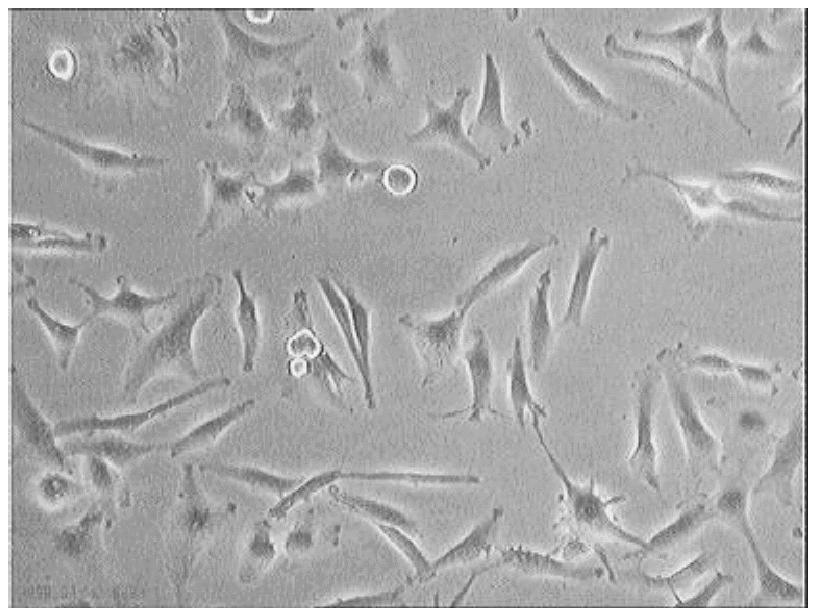,height=2.7in}}
\label{ha2x}
\end{figure}

\begin{figure}
\centerline{\psfig{figure=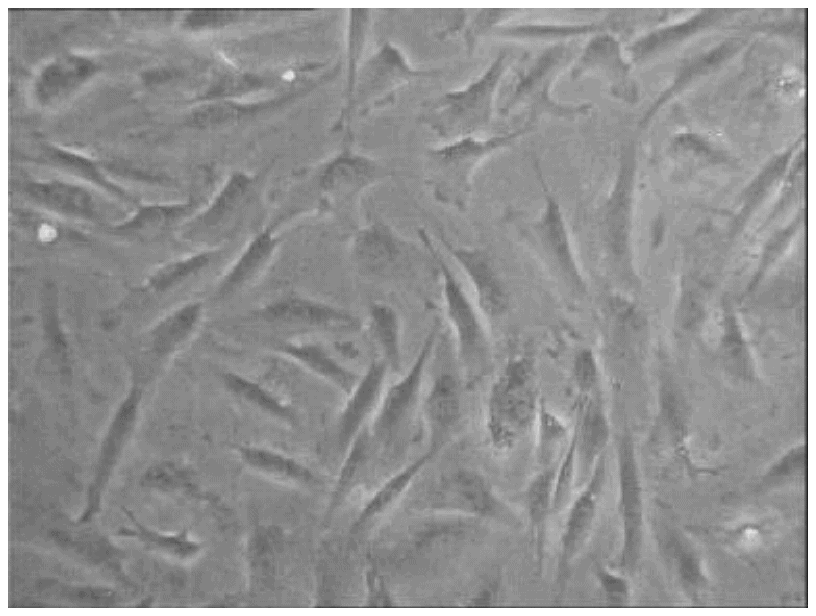,height=2.7in}}
\label{hb3x}
\end{figure}

\begin{figure}
\centerline{\psfig{figure=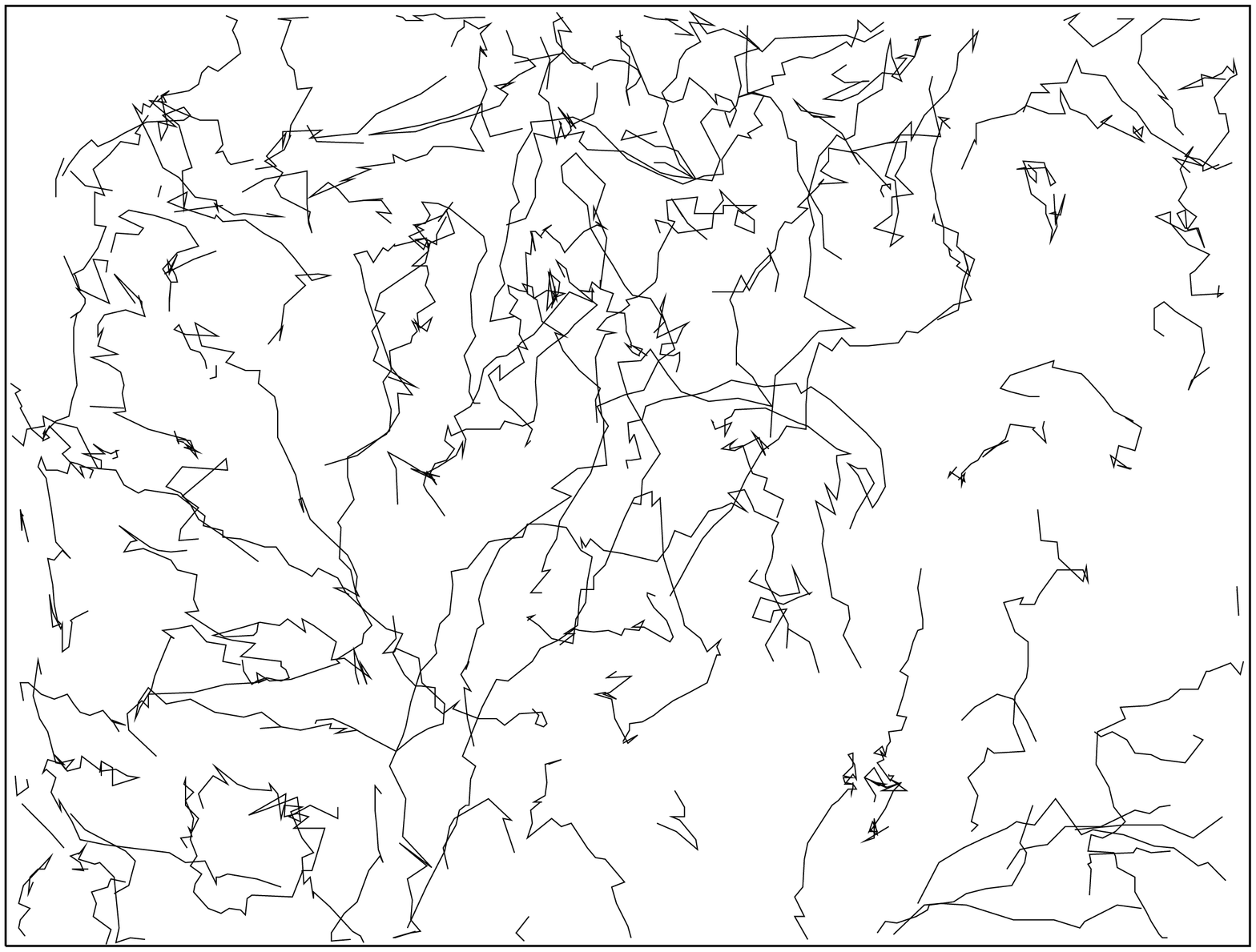,height=3.6in,angle=-90}}
\end{figure}

\begin{figure}
\centerline{\psfig{figure=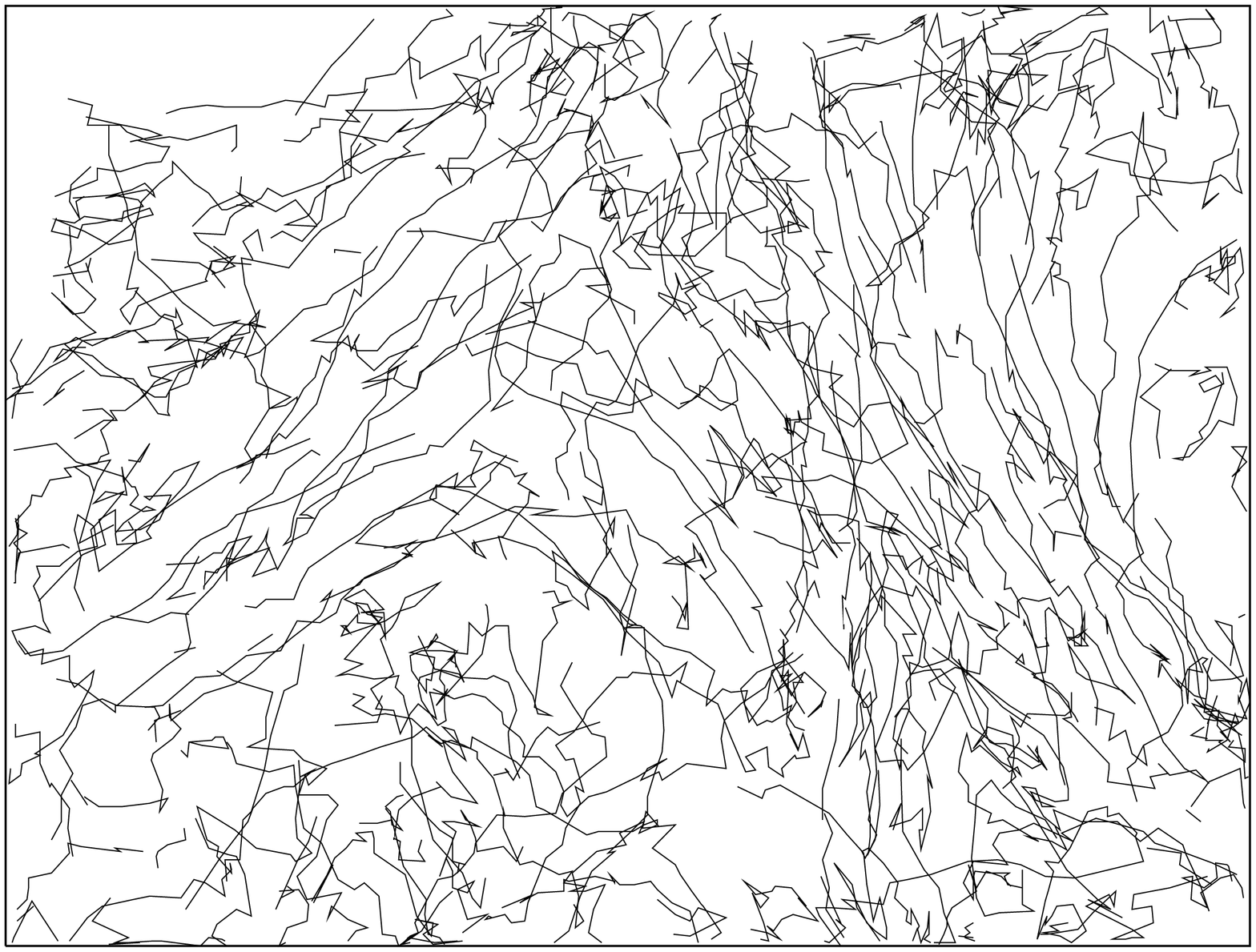,height=3.6in,angle=-90}}
\label{pathx}
\end{figure}

\begin{figure}
\centerline{\psfig{figure=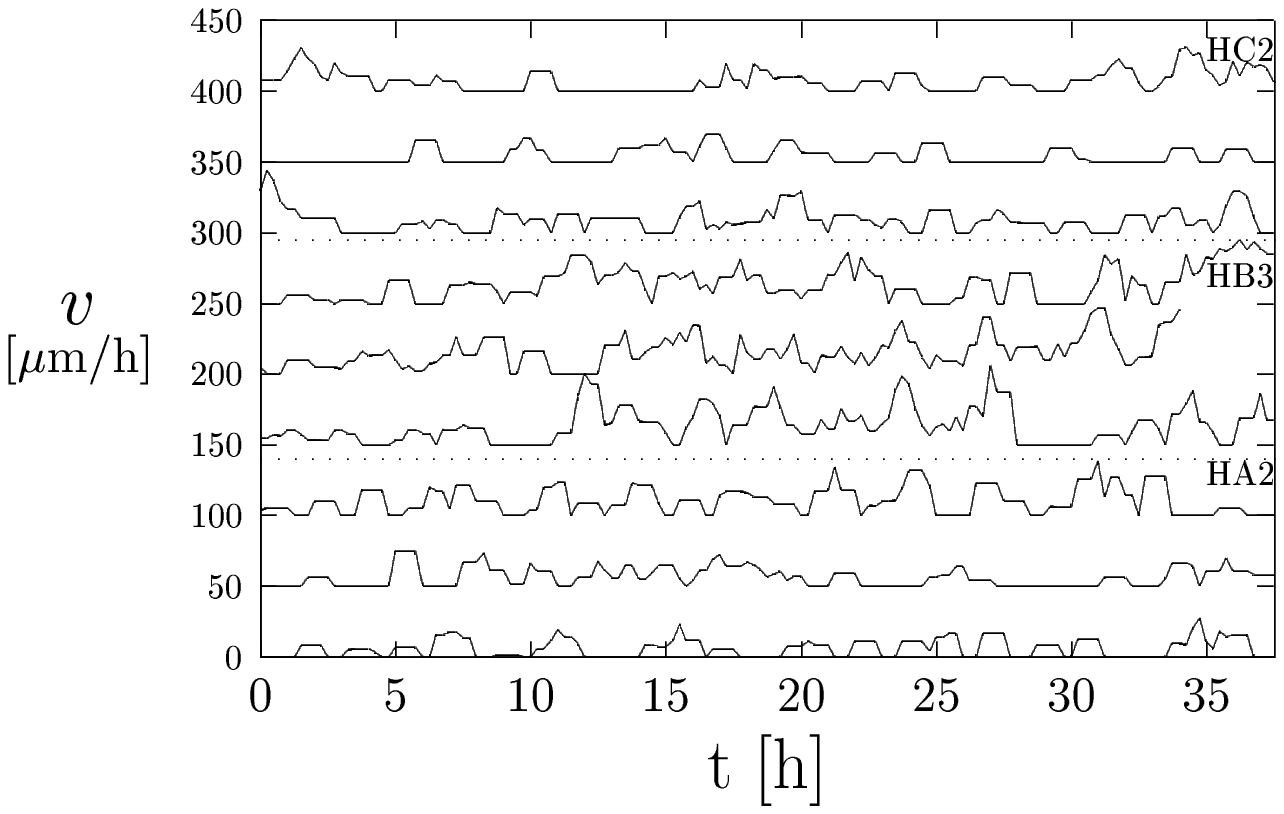}}
\label{fluctx}
\end{figure}

\begin{figure}
\centerline{\psfig{figure=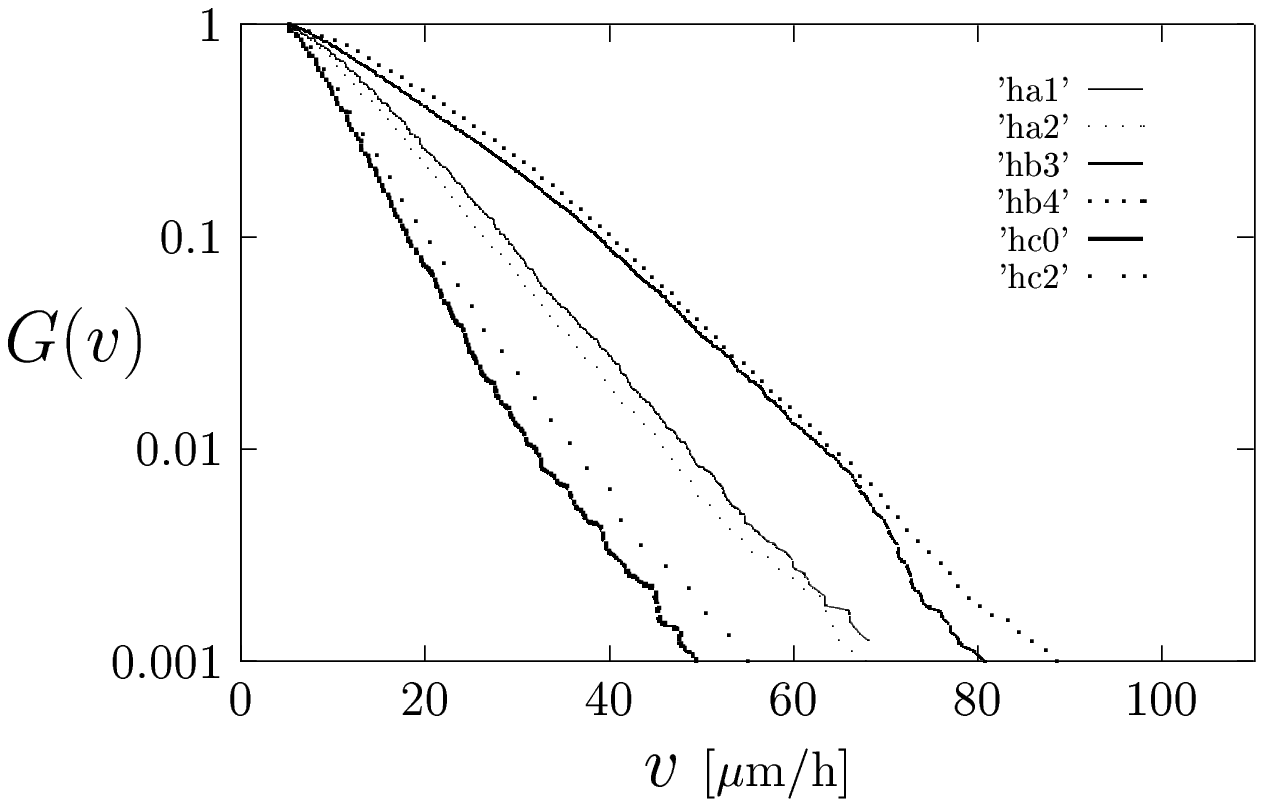}}
\label{veldisx}
\end{figure}

\begin{figure}
\centerline{\psfig{figure=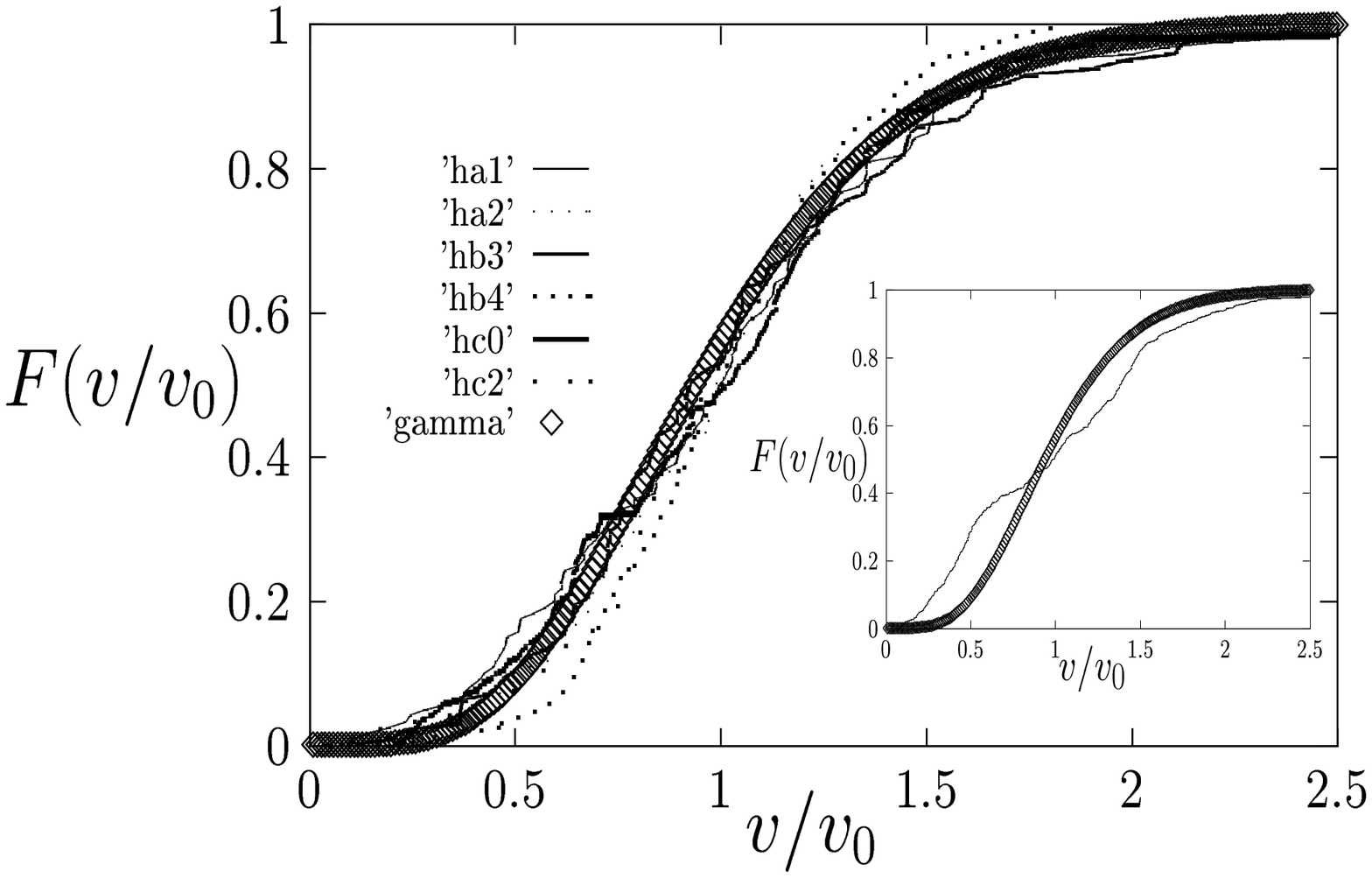}}
\label{gammax}
\end{figure}



\begin{thebibliography}{99}

\bibitem{Bern7} Bernstein JJ, Goldberg WJ, Laws ER Jr: Human malignant 
astrocytoma xenogtrafts migrate in rat brain: a model for central nervous
system cancer research {\bf J Neurosci Res 22}:134-143, 1989

\bibitem{big} Bigner DD et al.: Heterogeneity of Genotypic and Phenotypic 
Characteristics of Fifteen Permanent Cell Lines Derived from Human Gliomas 
{\bf J Neuropathol Exp Neurol 40}: 201-227, 1981

\bibitem{nature98} Cahill DP et al.: Mutations of mitotic checkpoint
genes in human cancers {\bf Nature 392}:300-333, 1998

\bibitem{rda} Chicoine MR, Silbergeld DL: Assessment of brain
tumor cell motility in vivo and in vitro. {\bf J Neurosurgery 82}:
615-622, 1995

\bibitem{rev} Chicoine MR, Silbergeld DL: Mitogens as motogens
{\bf J Neurooncol 35}: 249-257, 1997

\bibitem{rda2} Chicoine MR, Silbergeld DL: The in vitro motility of human 
gliomas increases with increasing grade of malignancy. {\bf Cancer 75}: 
2904-9, 1995

\bibitem{cza1}  Czir\'ok A, Schlett K, Madar\'asz E, Vicsek T: Exponential 
Distribution of Locomotion Activity in Cell Cultures {\bf Phys Rev Lett 81}:3038-3041, 1998

\bibitem{proli1} Deckert M, Reifenberger G, Wechsler W: Determination of 
proliferative potential of human brain tumors using the monoclonal antibody
Ki-67 {\bf J Cancer Res Clin Oncol 115}:179-188, 1989

\bibitem{dunn} Dunn GA and Brown AF: A Unified Approach to Analysing Cell 
Motility  {\bf J Cell Sci Suppl 8}: 81-102, 1987


\bibitem{ki67} Hoyt JW, Gown AM, Kim DK, Berger MS: Analysis of proliferative 
grade in glial neoplasms using antibodies to the Ki-67 defined antigen and 
PCNA in formalin fixed, deparaffinized tissues. {\bf J Neurooncol 24}:163-169, 
1995

\bibitem{neu83} Huang P, Allam A, Taghian A: Growth and metastatic behavior of
five human glioblastoma compared with nine other histological types of human 
tumor xenografts in SCID mouse.  {\bf J Neurosurgery 83}:308-315, 1995

\bibitem{who} Kleihues P, Burger PC, Scheithauer BW: Histological Typing of 
Tumours of the Central Nervous System {\bf World Health Organization, 
International Histological Classification of Tumours} Springer-Verlag: Berlin, 
Heidelberg, 1993

\bibitem{mozgas1} Lauffenburger DA, Horowitz AF: Cell Migration: A Physically 
Integrated Molecular Process {\bf Cell 84}:359-369, 1996

\bibitem{giant1} M\"uller W, Slowik F, Firsching R, Afra D, Sanker P:
Contribution to the problem of giant cell astrocytomas. {\bf Neurosurg
Rev 10}:213-219, 1987

\bibitem{Naz} Nazzaro JM, Neuwelt EA: The role of surgery in the management 
of supratentorial intermediate and high-grade astrocytomas in adults. 
{\bf J Neurosurgery 73}:331-344, 1990  

\bibitem{brdu}  Onda K, Davis RL, Shibuya M, Wilson CB, Hoshino T: Correlation 
between the Bromodeoxyuridine Labeling Index and MIB-1 and Ki-67 Proliferating 
Cell Indices in Cerebral Gliomas. {\bf Cancer 74}: 1921-1926, 1994


\bibitem{gfap} Rutka TJ, Murakami M, Dirks PB: Role of glial filaments in cells
and tumors of glial origin: a review. {\bf J Neurosurgery 87}: 420-430, 1997

\bibitem{nat77} Shields R: Transition probability and the origin of variation
in the cell cycle. {\bf Nature 267}: 704-707, 1977

\bibitem{mass} Silbergeld DL, Rostomily RC, Alvord EC Jr: The cause of death 
in patients with glioblastoma is multifactoral: clinical factors and autopsy 
findings in 117 cases of supratentorial glioblastoma in adults. 
{\bf J Neurooncol 10}:179-185, 1991
 
\end{thebibliography}
\end{document}